\documentclass[conference, compsocconf]{IEEEtran}

\IEEEoverridecommandlockouts
\usepackage{cite}
\usepackage{amsmath,amssymb,amsfonts}
\usepackage{graphicx}
\usepackage[caption=false]{subfig}

\def\BibTeX{{\rm B\kern-.05em{\sc i\kern-.025em b}\kern-.08em
    T\kern-.1667em\lower.7ex\hbox{E}\kern-.125emX}}
\usepackage{url}

\usepackage{ifthen,mdframed,xcolor,ulem}
\usepackage{multirow}
\usepackage{makecell}
\usepackage{adjustbox}
\usepackage{wrapfig}
\usepackage{balance}
\usepackage[numbers]{natbib}

\usepackage[boxed]{algorithm2e}
\usepackage{listings}
\definecolor{eclipseBlue}{RGB}{42,0.0,255}
\definecolor{eclipseGreen}{RGB}{63,127,95}

\lstdefinelanguage{sml}{
  basicstyle=\footnotesize\ttfamily,
  captionpos=b,
  frame = bt,
  framexleftmargin=0pt,
  framexrightmargin=-10pt,
  tabsize=2,
  columns=fixed,
  breaklines=true,
  comment=[l]{//},
  morecomment=[s]{/*}{*/},
  commentstyle=\color{commentsColor}\ttfamily,
  morekeywords= {
    EQUAL, GREATER, LESS, NONE, SOME, abstraction, abstype, and, andalso, array, as, before, bool, case, char, datatype, do, else, end, eqtype, exception, exn, false, fn, fun, functor, handle, if, in, include, infix, infixr, int, let, list, local, nil, nonfix, not, o, of, op, open, option, orelse, overload, print, raise, real, rec, ref, sharing, sig, signature, string, struct, structure, substring, then, true, type, unit, val, vector, where, while, with, withtype, word, set, *, ->
  },
  morestring=[b]",
  morecomment=[s]{(*}{*)},
  stringstyle=\color{black},
  identifierstyle=\color{black},
  keywordstyle=\color{blue},
  commentstyle=\color{eclipseGreen}
}

\newboolean{showcomments}
\setboolean{showcomments}{true} 
\ifthenelse{\boolean{showcomments}}
{ \newcommand{\mynote}[3]{
    \fbox{\bfseries\sffamily\scriptsize#1}
    {\footnotesize$\blacktriangleright$\textsf{\color{#3}{#2}}$\blacktriangleleft$}}}
{ \newcommand{\mynote}[3]{}}

\def\BibTeX{{\rm B\kern-.05em{\sc i\kern-.025em b}\kern-.08em
    T\kern-.1667em\lower.7ex\hbox{E}\kern-.125emX}}

\newcommand{\hide}[1]{}

\graphicspath{ {./images/} }

\title{
A Risk-taking Broker Model to Optimise User Requests placement on On-demand and Contract VMs
\thanks{This work is funded by Royal Thai PhD scholarship and EPSRC EP/R010528/1.}}

\author{\IEEEauthorblockN{Chalee Boonprasop}
\IEEEauthorblockA{
University of St Andrews\\
St Andrews, UK\\
Email: cb330@st-andrews.ac.uk}
\and
\IEEEauthorblockN{Yuhui Lin}
\IEEEauthorblockA{
University of St Andrews\\
St Andrews, UK\\
Email: yl205@st-andrews.ac.uk}
\and
\IEEEauthorblockN{Adam Barker}
\IEEEauthorblockA{
University of St Andrews\\
St Andrews, UK\\
Email: adam.barker@st-andrews.ac.uk}
}

\begin{document}

\maketitle
\begin{abstract}
Cloud providers offer end-users various pricing schemes to allow them to tailor VMs to their needs, e.g., a pay-as-you-go billing scheme, called \textit{on-demand}, and a discounted contract scheme, called \textit{reserved instances}. This paper presents a cloud broker which offers users both the flexibility of on-demand instances and some level of discounts found in reserved instances. The broker employs a buy-low-and-sell-high strategy that places user requests into a resource pool of pre-purchased discounted cloud resources. By analysing user request time-series data, the broker takes a risk-oriented approach to dynamically adjust the resource pool.

This approach does not require a training process which is useful at processing the large data stream. 
The broker is evaluated with high-frequency real cloud datasets from Alibaba. The results show that the overall profit of the broker is close to the theoretical optimal scenario where user requests can be perfectly predicted.

\end{abstract}

\section{Introduction}
Cloud providers offer various pricing schemes for the same \textit{Virtual Machines}~(VM) instance with different lease options and prices. The price differences between each scheme can be quite significant, e.g., a reserved instance can offer up to 75\% discount compared to the on-demand price \cite{ec2pricing}.
The numbers of options give users the flexibility to tailor VM instances to the requirements of their applications, however, it also complicates the choice of VM, given that there have already been large numbers of VM configurations offered in the market. 

\begin{figure}[htbp]
\center
\includegraphics[width=\linewidth]{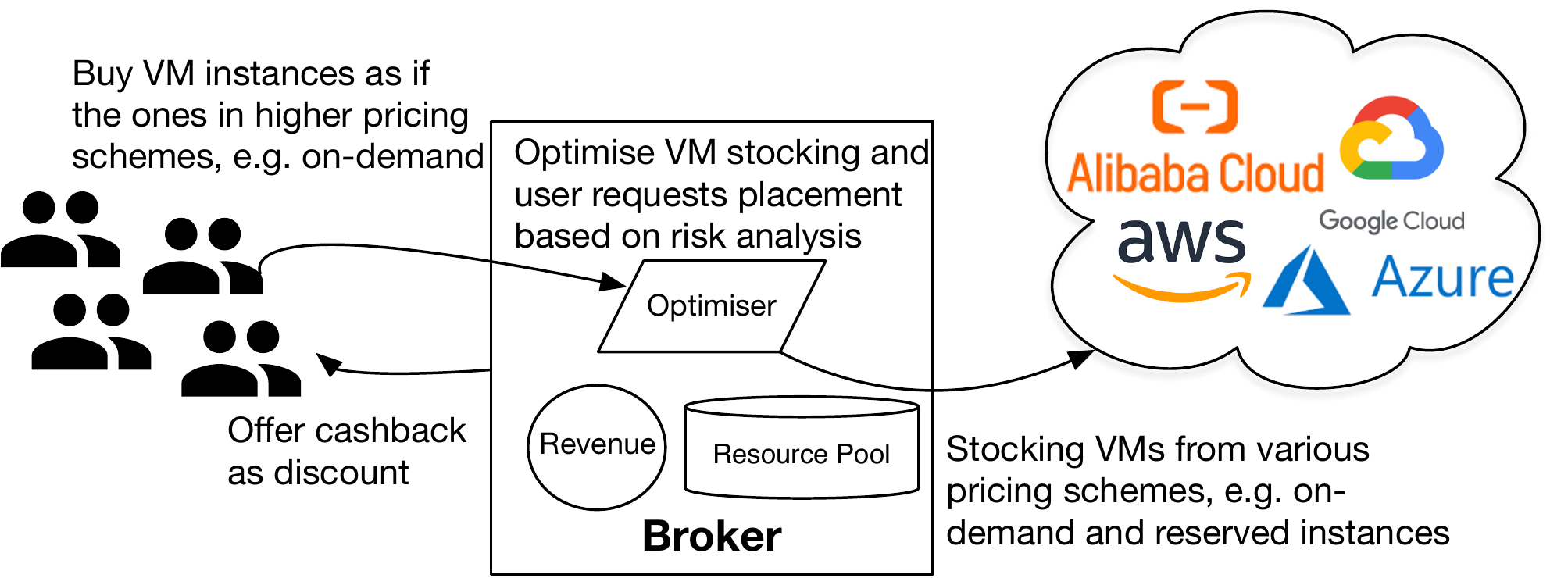}
 \caption{A broker model to simplify the choice of VMs from various pricing schemes. Users can both enjoy the flexibility as from on-demand instances and have discounts as from contract instance.} \label{fig: broker}
\end{figure}

Complexity creates opportunities. In this paper, we propose a broker model to simplify the choice of VMs from different price schemes, as shown in Figure~\ref{fig: broker}. The key to the strategy is to stock VM instance types from discounted contract pricing schemes in a \textit{resource pool} and then reselling them as VMs with a pay-as-you-go scheme to potential cloud users. 

Cloud users can then buy a VM instance as if the instance is from a higher pricing scheme to enjoy the flexibility of the instance whilst receiving cashback offered by the broker to reduce the cost. For the ease of presentation, we will take \textit{reserved instances}, which is a discounted scheme requiring commitments to a contract length, as a representative example for a lower pricing scheme, and \textit{ on-demand}, which is a flexible but more expensive pay-as-you-go scheme, for a higher pricing scheme. 

\begin{figure}[htbp]
\center
\includegraphics[width=0.7\linewidth]{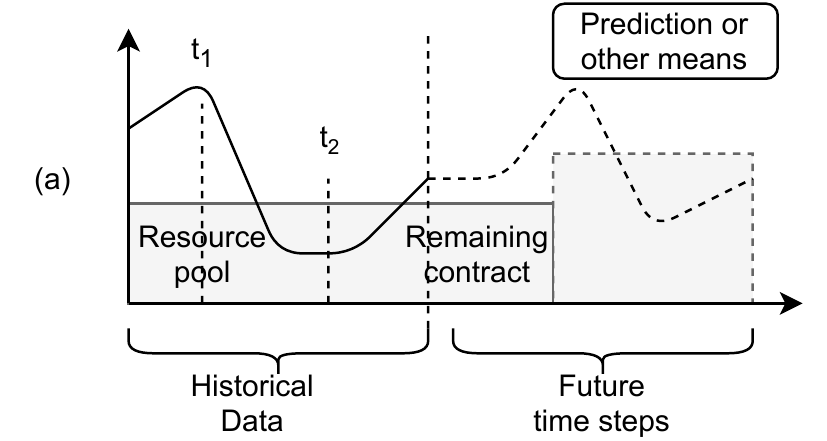}
\caption{Solving a time series data problem involves predicting future values. For example, at $t_1$ the current user orders exceed the resource pool size, some of the orders have to be offloaded to the on-demand instances; $t_2$ is a point where the resource pool is underutilised. Accuracy predictions are the key to reduce the cost and make the optimal decision on which scheme type of VMs the user order will be placed on.}
\label{fig:time_s_diff}
\end{figure}

The challenge of the strategy is how to optimise the stock of VM instance types from different pricing schemes according to the number of user requests. Figure \ref{fig:time_s_diff} illustrates the cases of over-stocking and under-stocking. 

Popular systems utilise the past data points to predict (time-series forecasting) the user requests values and then plan the size and composition of its resource pool \cite{jingmei,wang2017,wang2013}. The effectiveness of this method depends entirely on the assumption that past data is representative. However, the user requests in real-life are full of uncertainty, and it is non-trivial to justify whether a data set is representative or not.

Inspired by the risk-oriented trading strategy in the stock market~\cite{etula}, we take an alternative approach to drive the decision making with risk. We dynamically adjust the resource pool by evaluating the risks calculated from both the user requests data and the resource pool data. 
Our contributions are as follows:
\begin{enumerate}
    \item A VM resource optimisation that utilises risk analysis to dynamically adjust VM stocking level without assuming the underlying distribution of user requests.
    \item A generic broker system framework which can be extended to additional resources and risks for optimisation. 
\end{enumerate}

The paper is organised as follows. \S \ref{sect: rp opt} gives an overview of our broker and then explains our risk-analysis based decision-making process for placing user requests on different price schemes. \S \ref{sect: impl} gives a detailed account for our broker system. Evaluation of the broker system is discussed in \S \ref{sect: eval}, followed by the related work and future work in \S \ref{sect: related and future}. \S \ref{sect: concl} summarises this paper.

\section{Optimisation Resource Pool and User Request Placement using Risk Analysis}\label{sect: rp opt}
Our broker stocks reserved VMs instances in a resource pool but resells them as on-demand instances. The user experience is the same as buying on-demand instances directly from cloud providers, i.e., users can terminate VM at any time. When terminated, users can then get cashback as discounts. The cashback amount depends on the periods of VM that they use.

Internally, our broker places user requests in the VMs from the resource pool. In the case when there is no more VM available, the broker needs to decide on whether to stock reserved instances to fulfil the request or buy an on-demand instead. Such a decision-making process is at the centre of the broker. 

\begin{figure}[htbp]
\center
\begin{minipage}{\linewidth}
\begin{lstlisting}[language=sml,escapeinside={(*}{*)}, mathescape=true]
(*\includegraphics[width=0.8\linewidth]{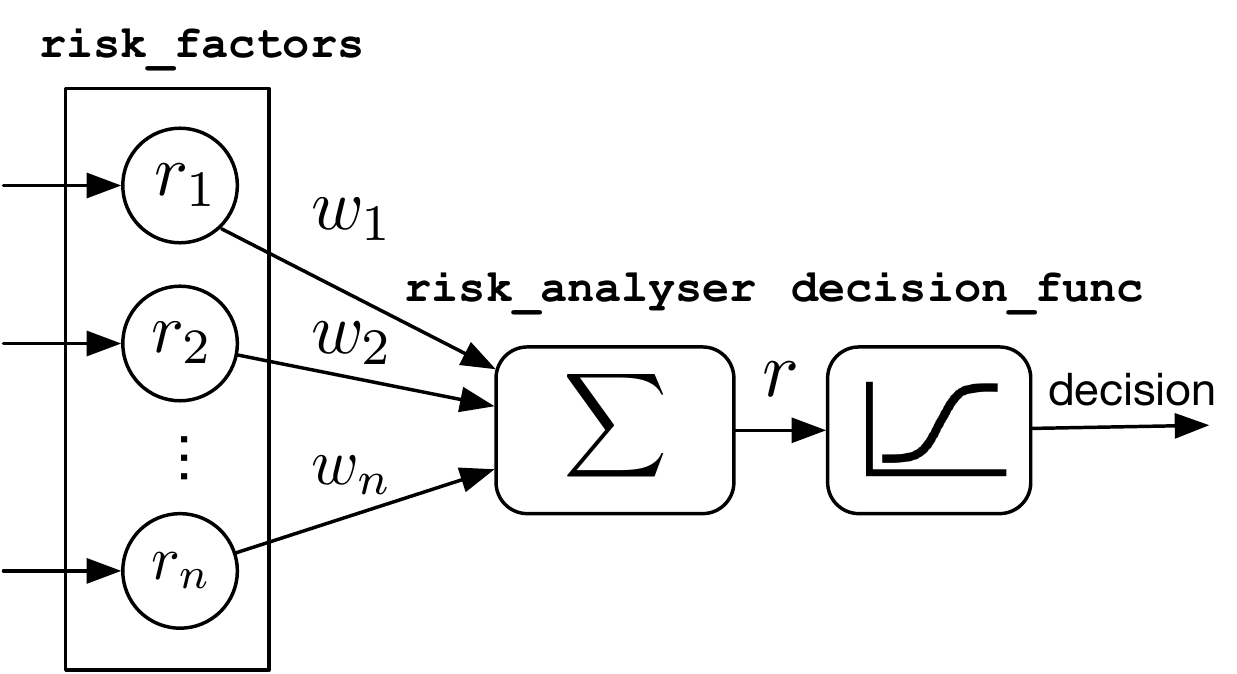}*)
(*\textbf{Input:}*)
// A set of quantitative risk factors 
risk_factors: (T -> $\mathbb{R})$ set
risk_analyser: $\mathbb{R}$ set -> $\mathbb{R}$
decision_func: $\mathbb{R}$ -> bool
(*\textbf{Output:}*)
// Define Optimisation with function composition
reinforce_dec = 
  decision_func $\circ$ risk_analyser $\circ$ map risk_factors
\end{lstlisting}
\end{minipage}
\caption{Risk-analysis based decision making process.}
\label{fig: risk analysis approach}
\end{figure}

We formulate the decision-making problem as a binary classifying function that taking quantitative risk factors as input. Figure \ref{fig: risk analysis approach} gives an overview of our risk-analysis based decision-making process, together with an abstract type definition of each part. 

There are three parts in the process: quantitative \textit{risk factors}, a \textit{risk analyser} and a \textit{decision function}. 
\begin{itemize}
    \item Each risk factor turns the status of brokers as a quantitative risk measure. The broker status covers the perspectives from the view of user requests and reserved instance. In the abstract definition, a generic data type (\texttt{T}) is used to allows a function in \texttt{\small risk\_factor} to accept parameters to produce a quantitative result for a risk factor.
    \item \textit{Risk analyser} normalises and assigns weights to each risk before aggregating them. There is also an additional risk-taking adjustment based on the current revenue level. The output $r$ is between [0,1].
    \item \textit{Decision function} takes the aggregated risk to decide whether to buy a reserved instance or an on-demand one to fulfil a user request.
\end{itemize}

In the rest of this section, we will give more details of each part of the process shown in Figure \ref{fig: risk analysis approach}.



\subsection{Quantitative risk factors}
We analyse the risks of stocking more reserved instances from the following aspects.

\textbf{Anomaly user requests}:
We should consider stocking more reserved instances if and only if the number of requests indicates an increasing trend in VM usage. In this case, we use a mean and standard deviation (\textit{mean-sd}) anomaly detection. Recall that our decision-making process is triggered by a periodic function which checks the pending queue, so we will use the same period as a time unit for our analysis. For the ease of presentation, we name the time of unit as $\mathcal{T}$ and define the user request rate as the number of the user requests over $\mathcal{T}$.

We quantify this factor by comparing the most recent user request rate to the mean of user request rate over the most recent 10\% length of a typical contract length of the reserved instance. For example, if the contract length is 10 months, the mean is calculated from the most recent 1 month. The method is also known as z-score anomaly detection \cite{killourhy2009comparing}.

The main point of this process is to reduce the number of outliers in the series data. The extreme values of outliers would artificially inflate the demand of the broker resulting in an unwanted escalation in cost. 

The threshold of the risk is 2 times the corresponding standard deviation, i.e., the range of anomaly requests is:$[mean, mean + 2 * sd]$

The quantitative risk factor of the anomaly requests is defined as:
\begin{equation}\label{eq: mean risk factor}
\begin{cases} 
      0 & r(t) < mean \\
      \frac{r(t) - mean}{2*sd} & mean\leq r(t)\leq mean+2*sd \\
      1 & r(t) > mean+2*sd 
\end{cases}
\end{equation}
where $r(t)$ is the time-series function at time $t$. To further improve the accuracy of risk analysis, on can introduce more anomaly detection to eliminate  other types of outliers. We will discuss possible directions in future work.

\textbf{Total numbers of reserved instances}:
The current number of reserved instances (the size of the broker inventory) is also a risk factor. If the current number is significantly higher than the average number, we should rate it as a substantially high risk. Similar to the anomaly request, here, we also take the mean of reserved instance size of the most recent 10\% of the typical contract length of the reserved instance for comparison. The function of calculating this risk factor is the same as \eqref{eq: mean risk factor}.

\begin{figure}[htbp]
  \centering  
  \includegraphics[width=0.6\linewidth]{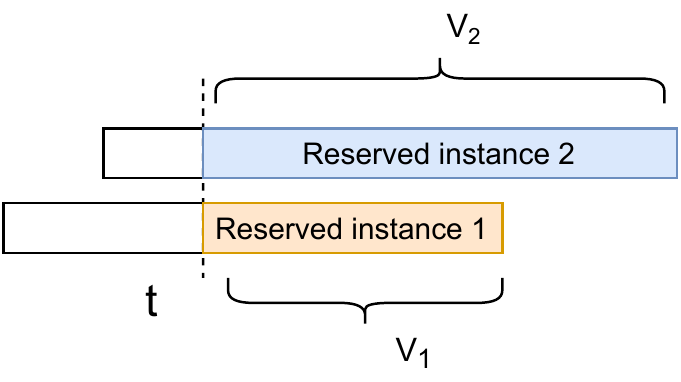}
  \vspace{-5mm}
  \caption{Using volume of the resource pools as a risk factor: $V_1 + V_2$}
  \label{fig:vol}
\end{figure}

\textbf{Volumes of reserved instances}: 
Another dimension of the risk of the current reserved instance stock is the volume, which is the remaining length of the contract. To illustrate, in the example shown in Figure~\ref{fig:vol}, the volume is $V_1 + V_2$.

The function to calculate the risk factor of volume is:
$$
\frac{\sum_{i=1}^{N} vol_i}{N * len}
$$
where $vol_i$ is the current volumes of a reserved instance; $N$ is the total number of the instances, and $len$ is the length of the contract.

\subsection{Normalised linear risk analyser}
To combine the risk factors listed, we use a linear risk model to normalise our risk values. The impact of each risk factor on profitability cannot be quantified without pre-existing data. Hence, throughout the experiment we consider all risk factors to be of equal importance. Also, for the ease of computation in the decision making, we would like to keep the range of the sum of all risk factor within $[0,1]$. So, we assign a weight vector to normalise the range of each factor accordingly, i.e., $\vec{w}^\intercal * \vec{r}$, where $\vec{w}$ is the weight vector and $r$ is a vector of the risk factors. The output value is the sum of normalised risk factors.

\textbf{Risk-taking risk adjustment}:
We would also like to take the current revenue level as a `positive' risk factor to adjust the sum of all the risk factor. The intuition behind is to allow higher risk to get more profit. This is a trading strategy in the traditional marketplace, called a risk-taking strategy. Here, we adopt a trading strategy in our resource pool adjustment. We compare the current revenue in the most recent period of $\mathcal{T}$ with the one in the period before. If the revenue rate is higher, we allow the broker to take more risks by reducing the risk factor by a fixed amount of 0.05. Similarly, it increases 0.05 when the revenue rate is lower. The number is arbitrary chosen to be 5\% of the maximum risk value. The main reason for allowing risk adjustment based on the money on hand is to identify the effect of opportunity cost and profitability of a broker. Balancing the amount of cash-on-hand is also one of the financial trade practices. In other words, cash is one of our soft risk factors. Note that the range of the risk factor after adjustment remain [0,1].

\subsection{Decision-making function}
\begin{figure*}[h]
\vspace{-5mm}
\begin{minipage}{0.32\textwidth}
\includegraphics[width=\linewidth]{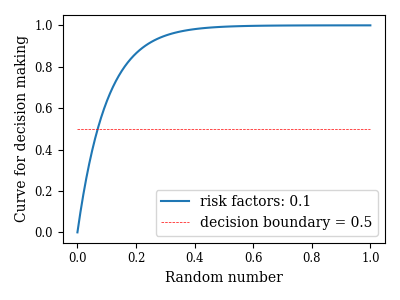}
\end{minipage}
\begin{minipage}{0.32\textwidth}
\includegraphics[width=\linewidth]{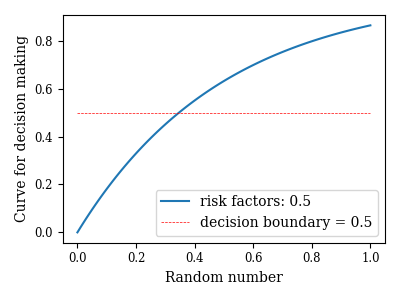}
\end{minipage}
\begin{minipage}{0.32\textwidth}
\includegraphics[width=\linewidth]{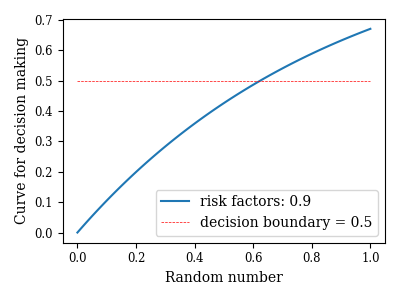}
\end{minipage}
\caption{Plots showing the decision curve when risk is low (0.1), medium (0.5) and high (0.9). x-axis is a random number generated between [0,1] with equal probability. When $y \leq 0.5$, the broker will create a reserved instance. The length of the arc curve represents the corresponding likelihood. A guideline of $y=0.5$ is provided for reference.}
\label{fig: decision}
\end{figure*}

Decision making is essentially a predicate which takes some of the quantitative risk factors to produce a boolean value for decision. In our case, we would like the decision function to satisfy the following requirements:
\begin{description}\small
    \item [(i)] Being able to take any value in the range of a sum of the risk factors, i.e. [0,1]
    \item [(ii)] The likelihood of creating a reserved instance changes continuously with the value of the sum of risk factors.
\end{description}
A non-deterministic function is used to satisfy the requirements, i.e.,
\begin{equation}\label{eq: deci}
    S(r) =
    \begin{cases} 
      1 & r = 0 \\
      1 - e^{-\frac{rng(0,1)}{r}} & 0 < r \leq 1 \\
\end{cases}
\end{equation}
where $rng(0,1)$ picks a random number between 0 and 1 with equal chance. When the output value is less and equal than 0.5, the broker will create a reserved instance to accommodate pending user requests, otherwise, an on-domain instance is created. With this function, the higher the risk is, the lower the chance of creating a reserved instance becomes. Also, the broker is strategically more inclined to allocate a reserved instance because the overall likelihood of getting a value below 0.5 is higher than the value above 0.5. Figure~\ref{fig: decision} shows three cases when risk is low (0.1), medium (0.5) and high (0.9). The randomness introduced into the system is inspired by the mutation algorithm. 

To decide for each pending user request, the broker will first compute the sum of the normalised each risk factors, and then generates the likelihood of each decision with a curve function, and finally `rolls a dice' to get a decision.

We have presented all the parts for our risk-analysis based approach to optimise resource pool and user request placement in the broker. A summarised pseudocode for our approach is shown in Figure~\ref{fig: risk opt}.

\begin{figure}[htbp]
\begin{lstlisting}[language=sml,escapeinside={(*}{*)}, mathescape=true]
(*\textbf{Input:}*)
// Functions to compute current risk factors 
risk_factors = {$\mathcal{F}_{anomaly\_rqst}, \mathcal{F}_{vm\_num},$
$\qquad \qquad \qquad \qquad \qquad \mathcal{F}_{vm\_vol}, \mathcal{F}_{profit}$}
// A linear risk model 
risk_model = $\lambda\vec{w}, \vec{risks}: \vec{w}^\intercal * \vec{risks}$
// A non-determined predicate to make decision 
decision_func = $\lambda r: 1 - e^{-\frac{rng(0,1)}{r}}$
// Weights to balance and normalise risk factors
$\vec{w}_{risks}$
/* Data from the broker: request db, resource pool and pending request queue */
rqst_db, rsrc_pool, rqst_srv_queue
(*\textbf{Output:}*)
/* Decision for whether to create a new reserved instance or an on-demand instance for each pending request */
decisions: bool list 
(*\textbf{Procedure:}*)
decisions = []
$\vec{risks}$ = [$\mathcal{F}_{anomaly\_rqst}(rqst\_db), \mathcal{F}_{vm\_num}(rsrc\_pool),$
$\qquad\qquad\;\mathcal{F}_{vm\_vol}(rsrc\_pool), \mathcal{F}_{profit}(current)$]
(*\texttt{\textbf{for}}*) $\forall$ rqst $\in$ rqst_srv_queue (*\texttt{\textbf{do}}*)
  x = decision_func($\vec{risks}$, $\vec{w}_{risks}$)
  decisions.append(x)
(*\texttt{\textbf{end}}*)
\end{lstlisting}
\caption{Detailed specification of the decision making process with risk analysis}
\label{fig: risk opt}
\end{figure}

In the next section, we will give a detailed design and implementation of our broker.

\section{Detailed Specification \& Implementation} \label{sect: impl}
To adopt the pricing strategy, we develop a broker system which consists of 6 main components: a {user request scheduler}, a {request database}, a {pending serving queue}, a {resource pool} and an {optimiser}. 
\begin{figure}[htbp]
  \centering
  \begin{minipage}{\linewidth}
  \includegraphics[width=\linewidth]{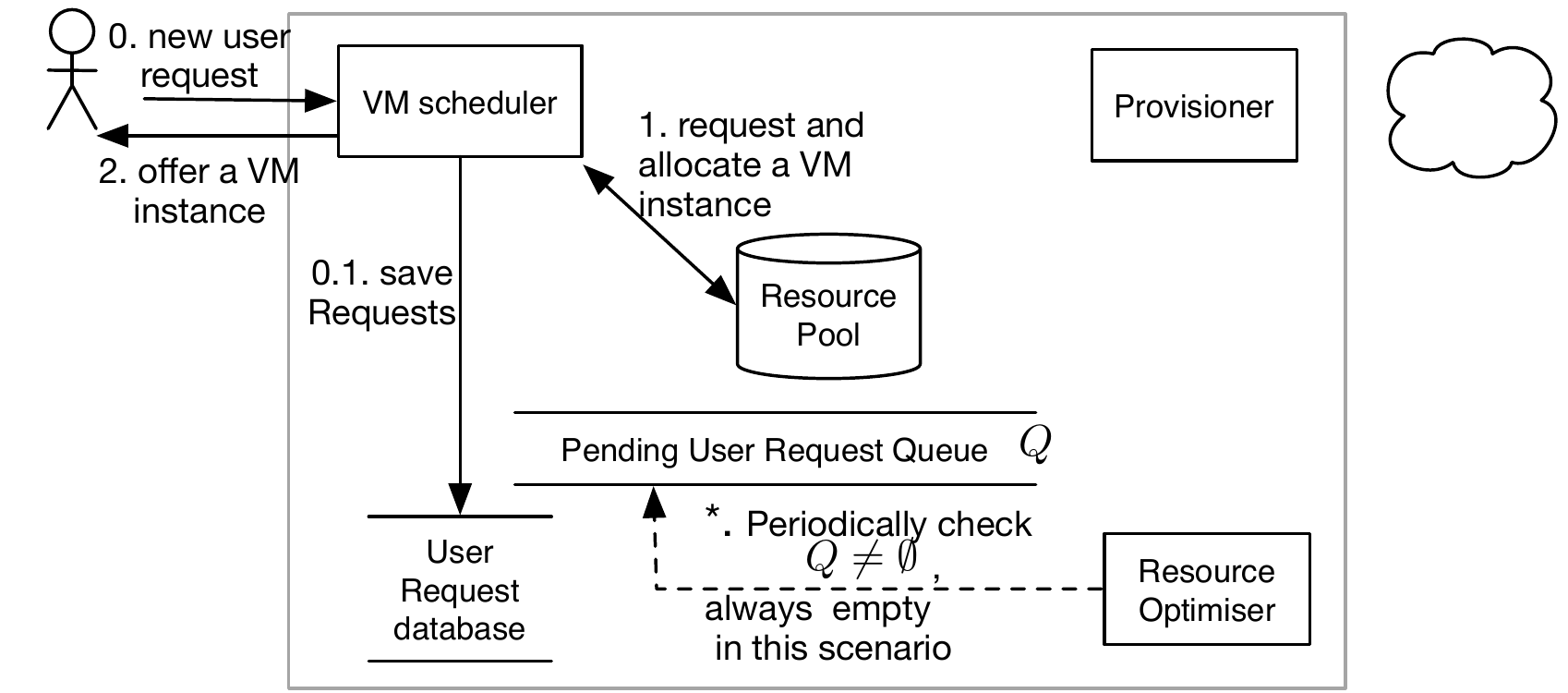}
  {\footnotesize(A) The broker will allocate a reserved instance directly from the resource pool to accommodate a user request when a reserved instance is available.}
  \end{minipage}\\
 \vspace{2mm}
  \begin{minipage}{\linewidth}
  \includegraphics[width=\linewidth]{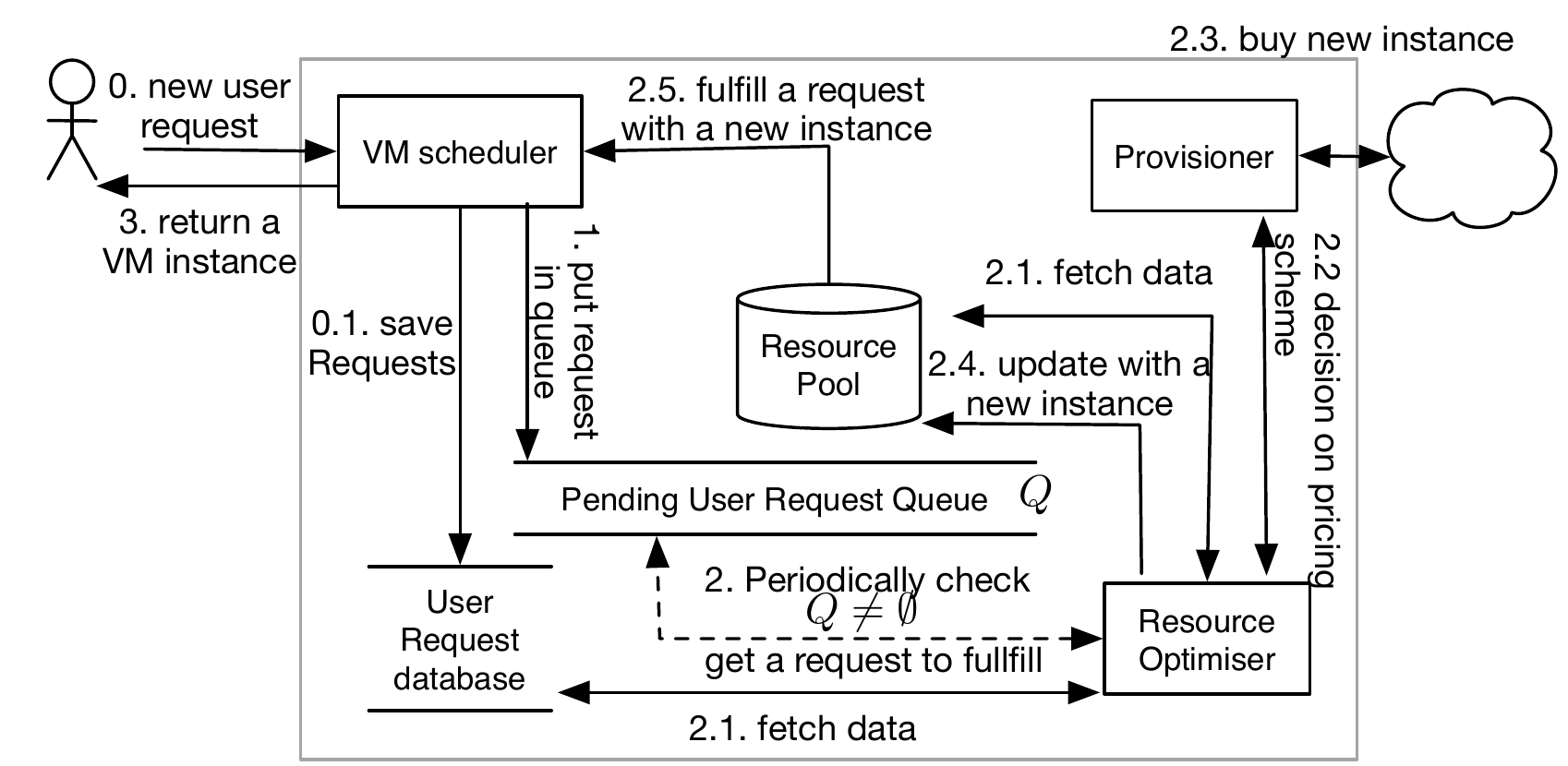}
  {\linespread{0.5}\footnotesize (B) The broker will create either an on-demand or a reserved instance to accommodate a user request  on when no reserved instance is available.}
  \end{minipage}
  \vspace{1mm}
  \caption{Workflows illustrating the broker's responses to a user request  when a VM is available (A) and when a no VM is available (B).}
  \label{fig:pricing strategy}
\end{figure}
Figure~\ref{fig:pricing strategy} shows different workflows of accommodating a user request. For the ease of clarification, we provide the following high-level abstract type definition of the broker and its component to walk through the functionality of the system:

\begin{figure}[htbp]
  \centering
\begin{lstlisting}[language=sml,escapeinside={(*}{*)}]
broker = (
 rqst_db: RqstDB_T, 
 rqst_srv_queue: RqstQ_T,
 rsrc_pool: RP_T,
 rqst_schdlr: Rqst_T * (RqstDB_T * RqstQ_T * RP_T) -> RqstDB_T * RqstQ_T * RP_T, 
 optimiser: RqstQ_T * RP_T -> RP_T,
 provisioner = (
  create_vm: VM_T * PrcSchm_T -> Rsrc_T,
  destroy_vm: Rsrc_T -> unit
 )
)
// Type Price Scheme 
type PrcSchm_T = Revered_T | OnDemand_T
// Type for User Requests 
type Rqst_T = UID * VM_T * CREATION | TERMINATION
// Type for the User Request DB
type RqstDB_T = (Rqst_T * time) set
// Type for the Request Queue 
type RqstQ_T = Rqst_T list
// Type for VM Resource
type Rsrc_T = RID * VM_T * PrcSchm_T
// Type for the Resrouce Pool 
type RP_T = Rsrc_T -> Rqst_T
\end{lstlisting}
\caption{Abstract type definition and specification for the broker model}
\label{fig: broker sys def}
\end{figure}

\noindent where \texttt{\footnotesize v: T} denotes a variable \texttt{\footnotesize v} with type \texttt{\footnotesize T}; \texttt{\footnotesize t\_1~*~...~*~t\_n} denotes the type of a n-dimensional tuple; and \texttt{\footnotesize X -> Y} denotes a function type. 

A broker is defined as a tuple of the 6 main components: a {request database} (\texttt{\footnotesize rqst\_db}) stores user requests as time-series data; a {resource pool}~(\texttt{\footnotesize rsrc\_pool}) maintains a collections of VM instances with different pricing schemes, as well as the occupying relationship between the VM instance and the user requests; a {pending serving queue}~(\texttt{\footnotesize rqst\_srv\_queue}) keeps track of the user requests that are yet to be fulfilled due to  lacks of reserved instances in the pool; a {user request scheduler}~(\texttt{\footnotesize rqst\_schdlr}) is a function to coordinate other components for request fulfilment depending on the availability of reserved instances; an {optimiser}~(\texttt{\footnotesize optimiser}) is an adjustment function of the stock of VMs in the resource pool in order to fulfil the requests in the pending queue; and a {provisioner}~(\texttt{\footnotesize provisioner}) is an agent to communicate with the cloud providers to create and terminate VM instances. 

Upon the arrival of each user request, the scheduler responds to the request by assigning/releasing the binding between the user request and the VMs in the pool. In the case of requesting to create a VM and the request cannot be fulfilled by the resource pool, the scheduler will put the request in the pending queue. The optimiser periodically checks if there are pending requests. If so, the optimiser will then review the current stocking level in the resource pool and a period of most recent user requests, to evaluate the risk level. To fulfil the pending requests, the optimiser can choose to either stock more reserved instances, if the risk is low, or buy on-domain instances in the case of high risk. The details of the process of risk analysis will be presented in the next section.

To keep the broker sustainable, another important aspect is the cashback model. 
Cashback is generated directly from the price difference between running cost and the income from reselling VMs. For example, if a user rents a VM for a $t_u$ unit duration and a broker made $P\%$ profit during the said duration. Then, the user is entitled to earn \textbf{at most} $P\%$ cashback of what originally spent.


The revenue of a broker is calculated from three main components, the user demands, broker running cost, and the cashback value. Each of the components affects the profitability of the broker system differently and certainly not trivial. If the broker aim is to maximise the profit, then it must be able to identify the correlation functions of each sub-component. The task is both challenging and dynamic. Therefore, in this work, we are only interested in the gross profit from the broker operation without diving too deep into the correlation functions. 

The gross profit margin ($\Psi$) from the time step $t_1$ to $t_2$ is calculated using the equation \ref{eq:profit}.
\vspace{-2mm}
\begin{equation}
\label{eq:profit}
\begin{split}
    \Psi(t_1, t_2) & = 100 * \frac{\rho(t_1, t_2) - \omega(t_1, t_2)}{\rho(t_1, t_2)}\\
    \rho(t_1, t_2) & = \sum_{i=0}^{n_{r}^{(t_1,t_2)}} (c_{Ond}u_{r}) \\
    \omega(t_1, t_2) &= \sum_{i=0}^{n_{Re}^{(t_1,t_2)}} c_{Re} + \sum_{i=0}^{n_{Ond}^{(t_1,t_2)}} (c_{Ond}u_{Ond_{i}})
\end{split}
\end{equation}

where $t_1$ and $t_2$ is the beginning and the end of the measured duration. $\rho(t_1, t_2)$ is the revenue of the broker during the same period with $c_Ond$ as the on-demand cost per unit time; $u$ as the time usage per request; and $n_r$ as the total number of requests. The operational cost of the broker ($\omega$) is comprised of the cost of the reserved instances and on-demand instances of the broker during the same period.

We have explained the details of our risk-based optimisation. In the next section, we will evaluate the broker using simulated user requests that are generated from Alibaba public cloud trace \cite{alibaba}. 

\section{Evaluation}\label{sect: eval}
To evaluate our broker model, we simulate a user request experiment environment using the Alibaba cloud datasets \cite{alibaba}. The profit level is used as our performance metrics with given user request data. In the rest of this section, we will present the details of the simulation environment setup and each broker models for comparison, followed by result analysis.

\subsection{Simulation environment}
In this experiment, we focus on evaluating the effectiveness of user requests placements for the same VM instance type from different price schemes. Therefore, we simplify the scenario settings by assuming that users only request for the same instance type. There are two pricing schemes, 3-month reserve instances and on-demand instances.

As each time unit elapses (which is one minute), our simulation environment feeds user requests to the experimental broker models, according to a given user request time-series data. 

We have prepared two sets of times series data. Each is generated from the Alibaba cloud server trace \cite{alibaba}. There are two versions of the trace: Alibaba 2017 and Alibaba 2018.
\begin{itemize}
    \item Alibaba 2017: Released in 2017, the trace lasted for 12 consecutive hours on 1300 machines. The trace includes a collocation of online services and batch workloads. 
    \item Alibaba 2018: Released in 2018, the trace lasted for 8 consecutive days on 4000 machines. The trace also contains the directed acyclic graph information of the batch workloads.
\end{itemize}

Both periods of the original datasets are not sufficiently long to evaluate the effect of user request placement for 3-month reserved instances. Therefore, we resample the user requests to be a 3-years' time-series data according to the method proposed by Moniz et al. \cite{moniz2017resampling}. Each request is a tuple of the following format:
$$ \small
\texttt{REQUEST\_ID * START\_TIME * TERMINATION\_TIME}
$$
Each resampled user request time series becomes an independent scenario for simulation. For the ease of presentation, we call the user request from Alibaba 2017 as \textbf{Dataset 1}, and the one from 2018 as \textbf{Dataset 2}. A summary is given in Table \ref{tab:simdatachange}.

\begin{table}[htbp]
\caption{Summary of User requests times series data for simulation} \label{tab:simdatachange} 
\centering
\vspace{-3mm}
 \resizebox{\linewidth}{!}{
 \begin{tabular}{c l l l}
    \hline
    \textbf{Scenario Ref} & \textbf{Source Data} & \textbf{Original Length} & \textbf{Resampled Length} \\
    \hline
    Dataset 1 & Alibaba 2017 & 12 hours & 3 years \\
    Dataset 2 & Alibaba 2018 & 8 days & 3 years \\
    \hline
 \end{tabular}
 }
\end{table}

\begin{table}[htbp]
\caption{Statistic description of both datasets}
 \label{tab:statdescribtion}
  \begin{center}
\begin{tabular}{c l l}
    \hline
    \textbf{Description} & \textbf{Dataset 1} & \textbf{Dataset 2} \\
    \hline
    Data Points & 1,298,775 & 7,324,831,146\\
    $\mu$/$\sigma$ & 77.08/96.98 & 49.12/370.43 \\
    Min/Max & 0/5,450 & 0/129,215 \\
    25 \% & 18 & 4\\
    50 \% & 47 & 14\\
    75 \% & 108 & 48\\
    \hline
 \end{tabular}
 \end{center}
\end{table}

Dataset 1 and 2 differs from each other in some key aspects. Table \ref{tab:statdescribtion} shows statistical description of the datasets. The size of Dataset 1 is smaller and observably less dispersing than dataset 2. Dataset 2 overall has higher volatility, which should have a direct impact on the profit of each system. We are expecting the result from Dataset 2 to be worse for the pure reserved strategy and a good challenge to the rest of the systems. 

\subsection{Broker models for comparison}
We set up two broker systems based on our broker models, i.e., \textit{No risk adjustment} and \textit{Risk-taking}. We also use three systems for comparison, where \textit{Pure reserved} and \textit{Best case} are for baselines and \textit{Auto-ARIMA} takes a typical approach using time series prediction. The details of each system are explained as below:
\begin{itemize}
    \item \textbf{Risk-taking}: The broker system is the completed version as we present in the \S \ref{sect: rp opt}. It takes all the risk factors into consideration as well as the risk-taking adjustment.
    \item \textbf{No risk adjustment}: This broker system is a part of Risk-taking broker system. It excludes the risk-taking adjustment when normalising risks.
    \item \textbf{Auto-ARIMA}: The future user requests are estimated using a time-series prediction technique, i.e., \textit{Auto-ARIMA} \cite{yermal2017application}. The exceed demands from the prediction are placed on the on-demand instance automatically. 
    \item \textbf{Pure reserved}: This is a naive system where all user requests are placed on reserved instances. The broker always stocks new reserved instances if there is any pending user request. We consider this system as a baseline for a lower boundary.
    \item \textbf{Best case}: The broker system can sneak peek the user requests information in the future and then plan the stock of VM instances accordingly. We consider this system as a baseline for the upper boundary.
    
\end{itemize}

\begin{table}[hbtp]
  \caption{Component usage in each broker strategies}
  \vspace{-3mm}
  \label{tab:strategy}
  \begin{center}
\begin{adjustbox}{width=0.48\textwidth}
\begin{tabular}{l|cc|ccc}
    \hline
    & \multicolumn{2}{c}{\textbf{Resource pool}} \vline& \multicolumn{3}{c}{\textbf{Optimisation}}\\\cline{2-6}
    \textbf{System} & \textbf{Reserved} & \textbf{On-demand} & \textbf{Risk} & \textbf{Adaptive Risk} & \textbf{Prediction}\\
    \hline
    Risk-taking & \checkmark & \checkmark&\checkmark &\checkmark &\\
    \hline
    No risk adjustment & \checkmark & \checkmark & \checkmark& &\\
    \hline
    Auto-ARIMA & \checkmark & \checkmark & & & \checkmark \\
    \hline
    Pure reserved & \checkmark & & & &\\
    \hline
    Best case* & \checkmark & \checkmark & & & Exact  \\
    \hline
 \end{tabular}
\end{adjustbox}
\end{center}
* The broker makes optimal decisions based on perfectly accurate prediction for the future user requests. 
\end{table}

Table \ref{tab:strategy} shows components in each system. All the brokers will have reserved instances as their main resource pool. Apart from the pure reserved, the rest of the systems use on-demand instances as a buffer when appropriate. The no risk adjustment strategy employs a fixed risk when making a decision, whereas the risk-taking can alter the risk level according to the cash-on-hand level. The Auto-ARIMA is an automatic variable adjustment time-series prediction model which place excessive demands from the prediction onto on-demand instances. Lastly, the best case is the system which produces the highest profit for the given data.

\subsection{Experimental results}
The results of the simulation show comparisons of accumulated quarterly profit. The calculation of the profit uses the data taken from Alibaba compute type pricing where the reserved instance gives 60\% static discount. And, the profit is calculated from Equation \ref{eq:profit}. Lastly, the simulation does not account for communication time between users, broker, and providers. 

\begin{figure}[htbp]
  \centering
  \includegraphics[width=\linewidth]{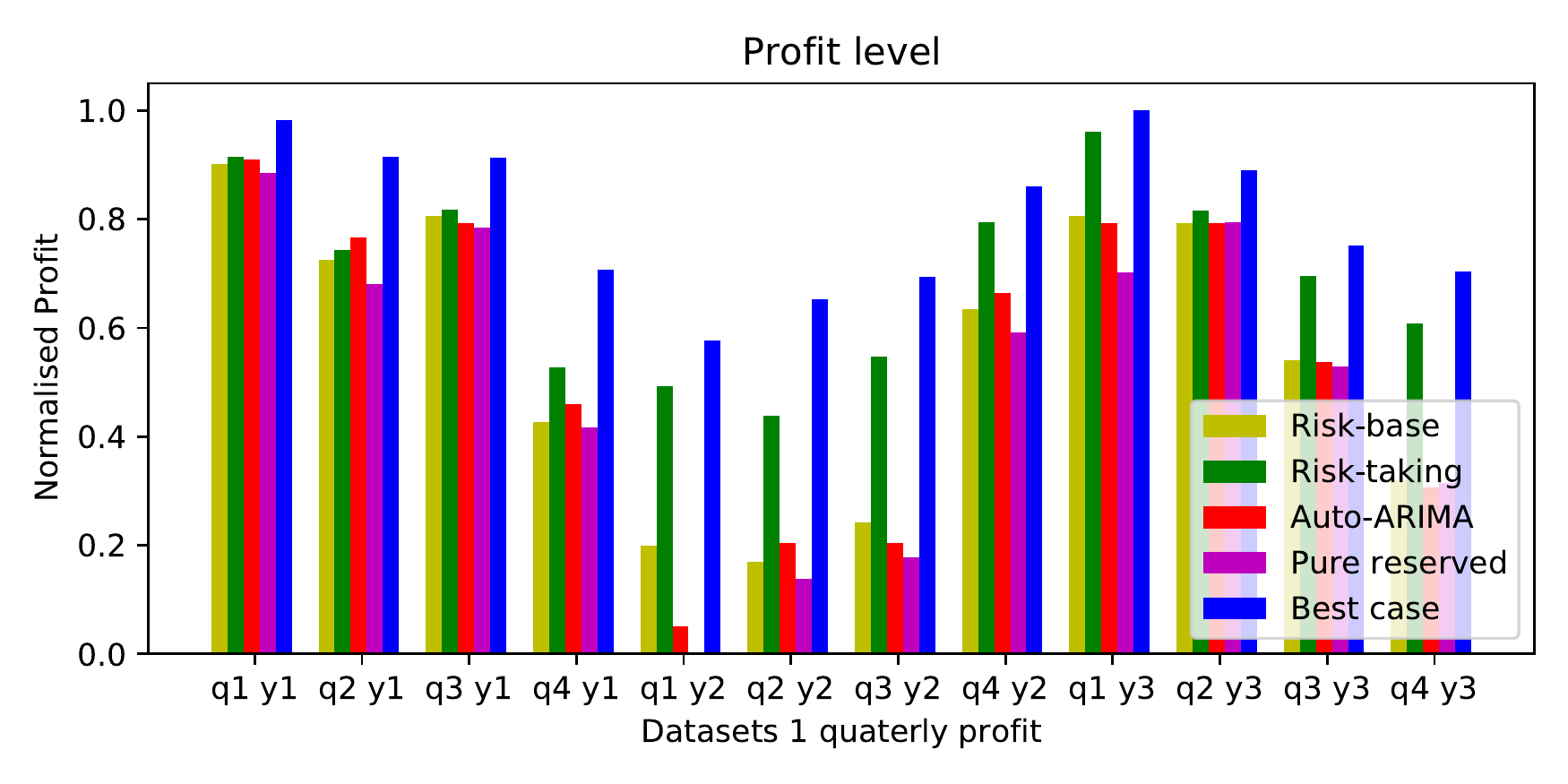}
  \caption{The graph shows a comparison of profits between each broker system of the input Dataset 1. The broker system components of each system are detailed in Table \ref{tab:strategy}. The input is divided into multiple parts of 4 months to better illustrate the profit level in each period. Both risk-based systems outperform the pure reserved and reach close to the theoretical maximised profit. The results are consistent throughout the data.}
  \label{fig:result1}
\end{figure}

Figure \ref{fig:result1} shows the profit of each broker system from Dataset 1. Empirically, both of our systems outperform the pure reserved strategy and Auto-ARIMA model for most parts. In some quarters, our risk-based produce profit levels comparable to the best-case scenario. The statistical break down of the result is shown in Table \ref{tab:summary}. In comparison to the best case, the overall profit of both risk systems has a higher high and higher low than competing systems. Strictly speaking, higher high and low means that it is more likely to earn a higher profit while less likely to lose money. 

\begin{table}[h]
  \caption{Comparative values from the base case}
  \label{tab:summary}
  \begin{center}
  \begin{tabular}{l c c}
    \hline
    \textbf{Systems} & \textbf{Highest profit(\%)} & \textbf{Lowest profit (\%)} \\
    \hline
    Risk-taking &96.06 & 65.85 \\
    No risk adjustment & 91.51 & 23.08 \\
    Auto-ARIMA & 92.50& 21.60\\
    Pure reserved & 89.85 & -4.42 \\
    \hline
\end{tabular}
\end{center}
    \vspace{-5mm}
\end{table}

\begin{figure}[h]
  \centering
  \includegraphics[width=\linewidth]{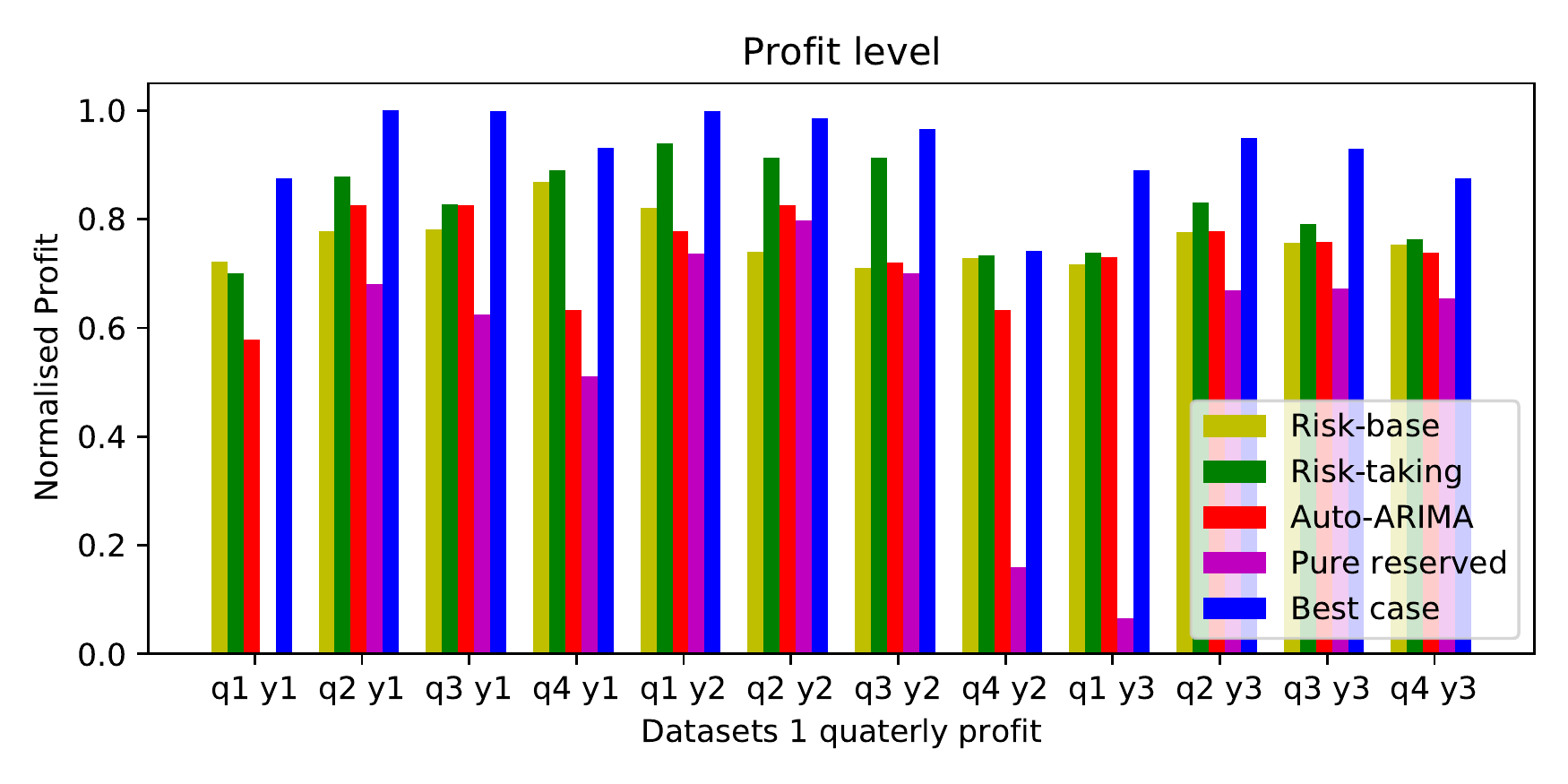}
  \caption{The graph illustrates normalised profit level of input dataset 2. In this dataset, the pure reserved instance struggles to return a profit. On the other hand, the risk-based system manages to stay close to the best case. With the inclusion of risk adjustment feedback, the profit level manages to edge closer to the actual best case values.}
  \label{fig:result2}
\end{figure}

\begin{figure}[h]
  \centering
  \includegraphics[width=\linewidth]{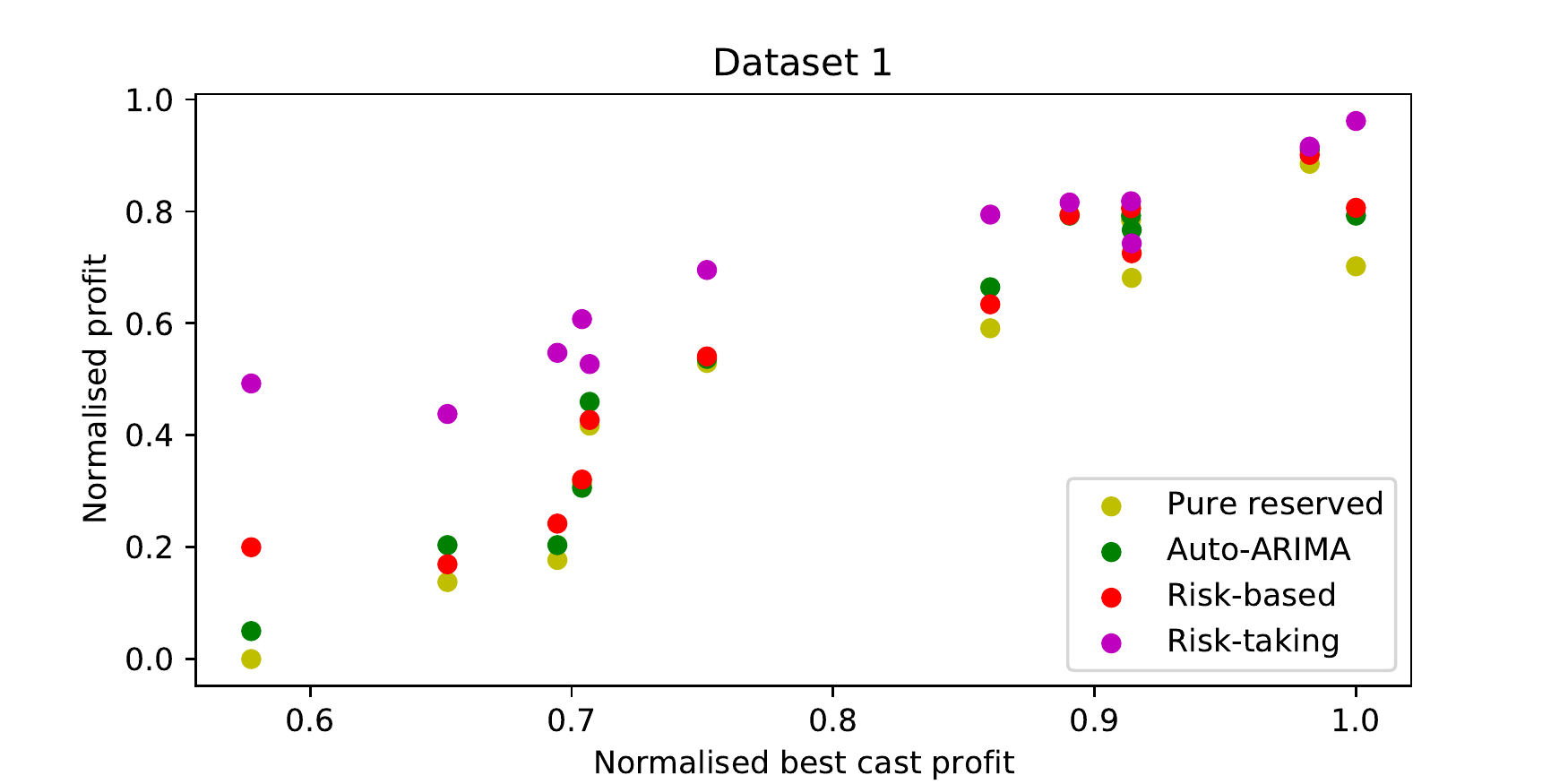}
  \includegraphics[width=\linewidth]{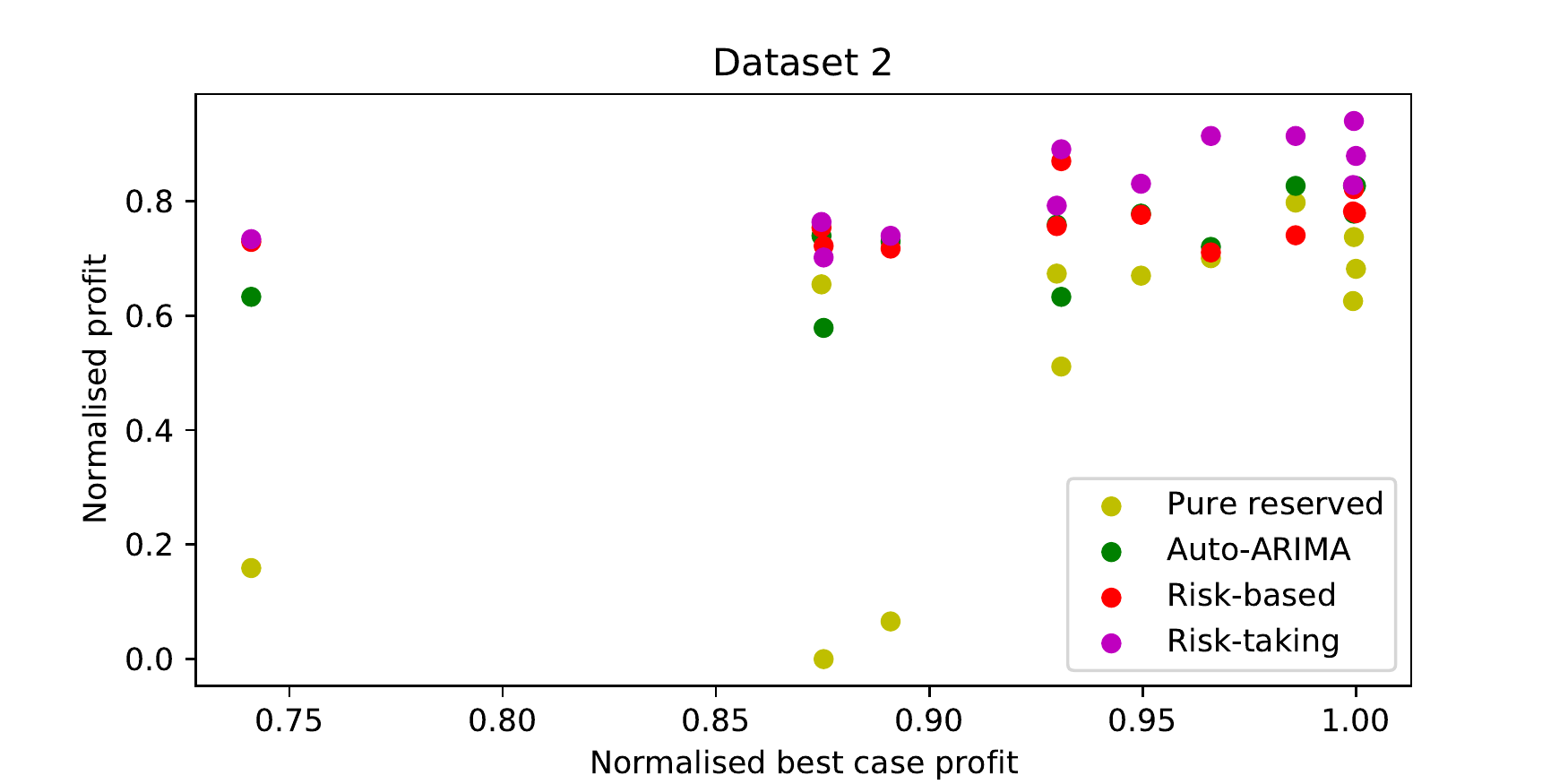}
  \caption{(TOP) A scatter plot of a normalised profit level of Dataset 1. The magnitude of the difference between each system is relatively similar throughout the system. Generally, we can see a pattern that the risk-taking outperform normal no risk adjustment system and Auto-ARIMA system for the majority of the period. (BOT) A scatter plot of a normalised profit level of Dataset 2. In Dataset 2, we can see that the Auto-ARIMA and no risk adjustment perform similarly while still trailing behind the risk-taking. Overall, the difference is larger than that of Dataset 1 result.}
  \label{fig:result2_2}
\end{figure}

We have normalised the profit level in a min-max normalisation fashion and compare the results from each system with the best case in Figure \ref{fig:result2_2}. All systems perform on a similar trend to the best case with our risk-based outperform the Auto-ARIMA in both Datasets. Without the offloading capability, the reserved only system vastly underperforms the rest of the systems. This is especially pronounced in Dataset 2 where data is more volatile. 

Additionally, we also experiment with the risk level such that the broker taking less risk to allow high utilisation in its resource pool. With the assumption that a system with high utilisation of the resource pool, the system is likely to have a higher profit. A sample of the first quarter of the second year (q1 y2) period in the Dataset 1 shows that higher average utilisation does not equate to higher profit. The broker employs less reserved instances in the resource pool; thus, it has to rely more on the on-demand instances which drive the cost higher. The issue is also known as opportunity cost. To make a good profit and avoid losing money, a broker has to strike a balance between the risk and reward, which is the theme of this work.

\begin{figure}[htbp]
  \centering
  \includegraphics[width=7cm]{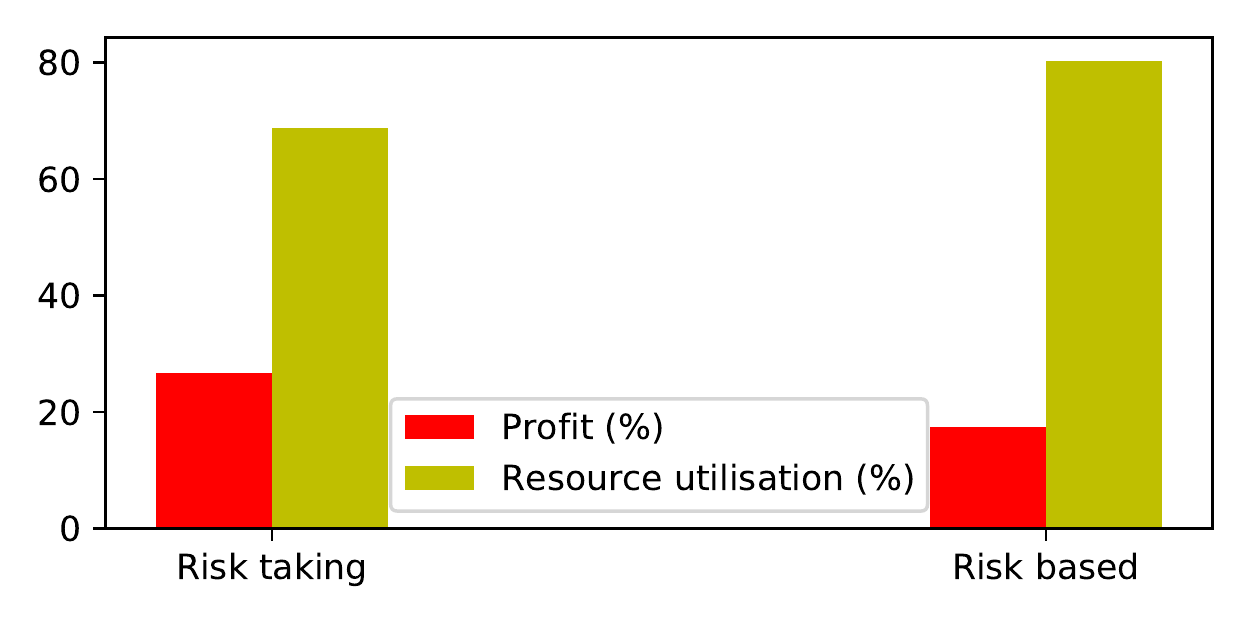}
  \caption{The graph shows the profit of the broker in the q1 y2 period in Dataset 1. From the average utilisation of both systems, the risk-based has a higher utilisation lower profit. A higher average reserved instance usage does not always translate to a higher profit.}
  \vspace{-3mm}
  \label{fig:result4}
\end{figure}

\begin{figure}[htbp]
  \centering
  \includegraphics[width=9cm]{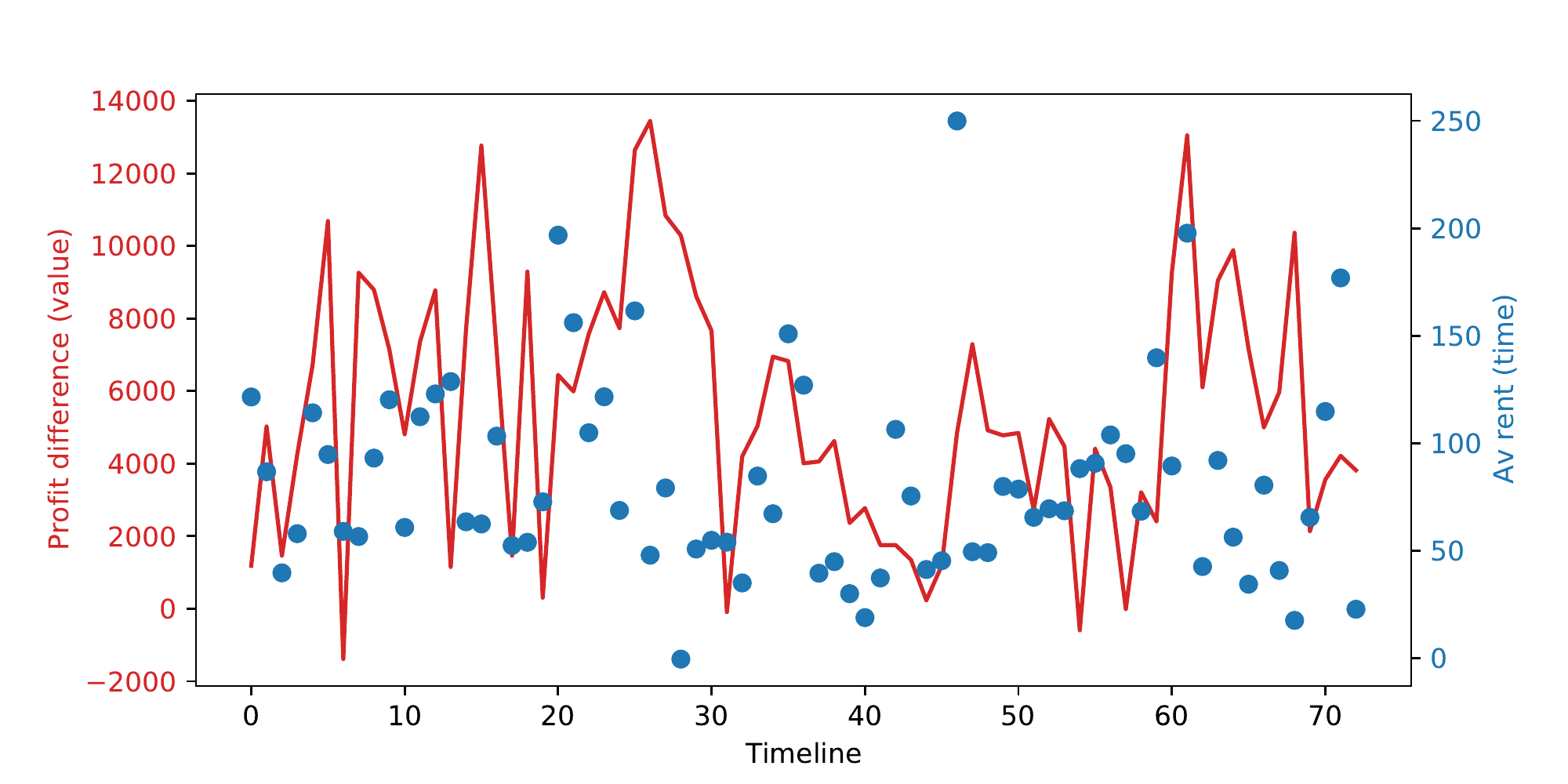}
  \caption{The graph of a profit difference between pure reserved and the best case and average usage time of users. A typical assumption for highly frequent small requests data is that it should suit the broker with more on-demand instances rather than the one that relies heavily on the reserved instances. However, from the graph, it does not appear to be the case.}
  \vspace{-3mm}
  \label{fig:timedif-avrent}
\end{figure}

\begin{table}[h]
  \caption{The correlation tests between profit difference of best case and pure reserved and average usage time of users.}
  \label{tab:timedifcorrelation}
  \begin{center}
  \begin{tabular}{c c c}
    \hline
    \textbf{Test} & \textbf{Correlation} & \textbf{P-value} \\    \hline
    Pearson & 0.11 & 0.34 \\
    Spearman & 0.09 & 0.45 \\
    Kendall & 0.06 & 0.46 \\
    \hline
\end{tabular}
\end{center}
    \vspace{-5mm}
\end{table}

Apart from the utilisation relation with profit, it may seem intuitive to assume that a short high-frequency type of usage would be suitable for a broker that utilised more on-demand instances. To investigate the claim, we overlay a profit difference and average cloud usage in Figure \ref{fig:timedif-avrent}. The correlation looks to be good in some time steps but bad in some others. The correlation tests shown in Table \ref{tab:timedifcorrelation} further confirm that there is an exceptionally weak correlation between the short frequency and profit differences in all three tests \cite{faul2009statistical}. Hence, it is unlikely that the average usage time affects the profit level of an on-demand bias broker. On the contrary, the risk factors level, as opposed to the usage time, shows a better correlation with the profit. The correlation valves from multiple correlation tests are shown in Table \ref{tab:timedifcorrelation-2}. The correlation is by no means a conclusion that the risk factors are the best indicator of profit. But it is a good signal that the system works.  

\begin{table}[h]
  \caption{The correlation tests between profit difference of best case and pure reserved and risk factors.}
  \label{tab:timedifcorrelation-2}
  \begin{center}
  \begin{tabular}{c c c}
    \hline
    \textbf{Test} & \textbf{Correlation} & \textbf{P-value} \\    \hline
    Pearson & 0.71 & 1.6e-12\\
    Spearman & 0.68 & 1.8e-11 \\
    Kendall & 0.57 & 7.8e-13 \\
    \hline
\end{tabular}
\end{center}
    \vspace{-5mm}
\end{table}

\begin{figure}
    \centering
    \includegraphics[width=\linewidth]{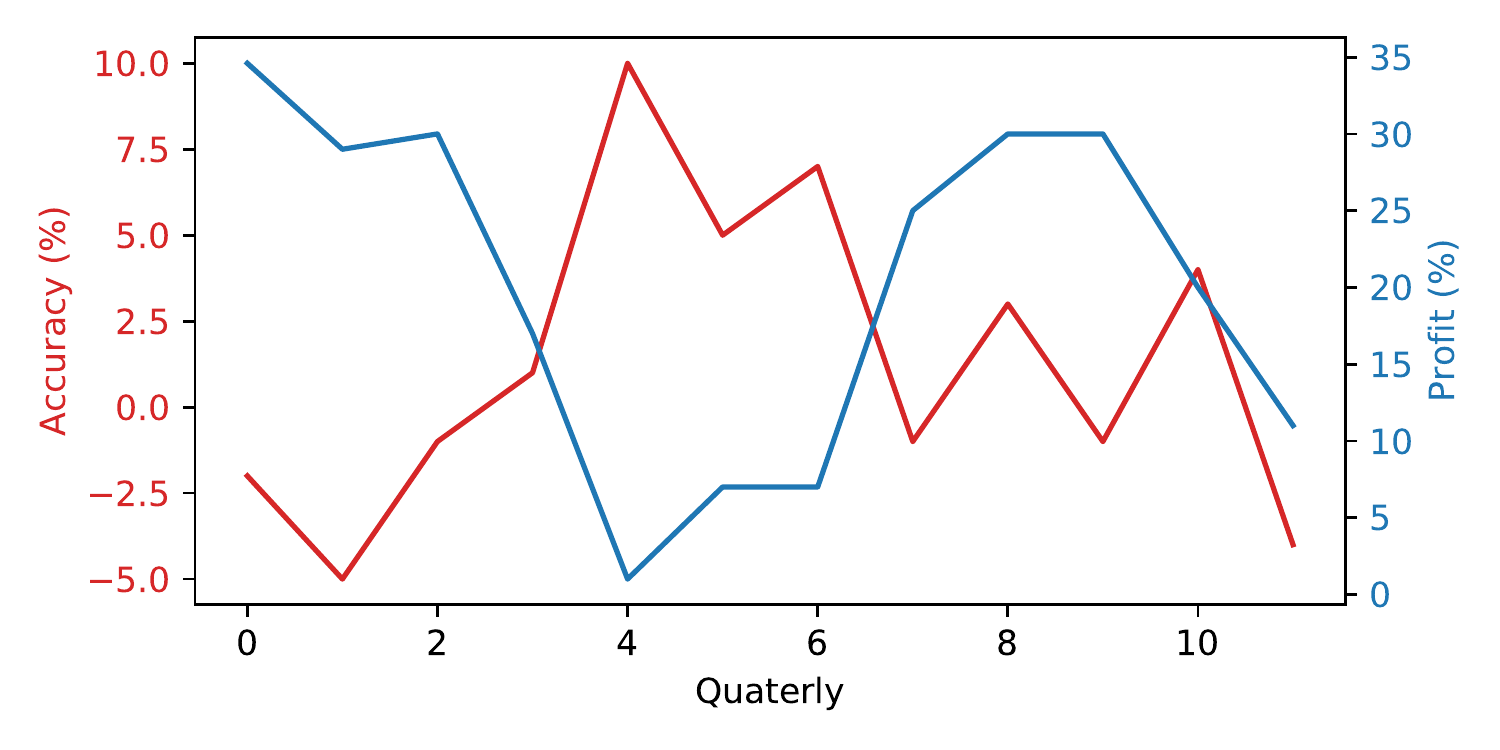}
    \caption{The graph compares percentage of over or under estimate the future values of the Auto-ARIMA (RED) and quarterly profit (BLUE). From the result, the profit of a broker seems to be negatively effect by the over-estimation rather than the under-estimation.}
    \label{fig:accuvsprof}
\end{figure}

The main concern in a cloud broker system that aims toward good profit is the low utilisation of its resource pool. The situation occurs when the system over-estimate the demand of its users and prepare larger than necessary resource pool. Figure \ref{fig:accuvsprof} shows the over-under estimation of Auto-ARIMA prediction and its respective profit. The overestimation (positive red number) causes the system to lose more money than the underestimation values. 

The solution to the problem could be to place some bias toward underestimation prediction. In our system, the bias toward these correlations to a profit is already indirectly built into the risk factors. With the consideration of the remaining time and size of the resource pool, our system can soften the effect of overestimation. 

If more conditions of a cloud broker are introduced into the system, the system can become complex and risk analysis must be reevaluated to keep up with the additional complication. Although, we cannot cover all of the broker scenarios with just three risk factors. We have shown empirically that our risk system still works even with a simple and small number of risk factors. The result is achieved without running an optimisation algorithm which takes considerably more time and an accurate user data regression. 

\section{Related work \& Future work}\label{sect: related and future}
Research on the profit maximisation of cloud computing has been performed using multiple approaches. We will discuss some of the related ones below. 

In the cloud computing space, the research on resource scheduling has been done on the basis that the scheduler knows both the starting and the termination time of the requests \cite{jingmei2, chaisiri, owen}. The starting requests of the broker are assumed in the form of probability distribution and the termination requests are attached to the starting requests. The method of their scheduling aims to maximise the utilisation of the reserved instances. The results are mathematically optimised. However, the assumption of the broker knowing the starting and termination time of user requests is not always guaranteed, and as a result, could potentially constrain the applicability in real life. 

Another approach in the profit-making of the cloud broker is also studied by Amit et al \cite{amit}. They considered QoS parameters in a profit aware model for the providers. The data communication model that reports the usage pattern and type of requests is used to increase the utilisation and profit of the provider. This is similar to our approach in terms of using the prediction pattern. However, the communication model considers the resource as a fixed cost - whereas our broker considers the resource as a varying cost. 

Additionally, job scheduling was also considered a profit maximising strategy. Shalmali, et al. claimed that a good job-scheduling model would increase the profit for cloud computing \cite{shalmali}. Similarly, a queuing model is used to maximise the profit for cloud computing \cite{cao2012optimal}. Both approaches toward maximising profits used QoS parameters with a derived probability density function of the service requests.

Profit maximising is also explored in the form of requests allocation on reserved instances. The profit is maximised based on the varied contracted lengths of the reserved instances and the assumed number of users' requests \cite{jingmei2, jingxiao}. The work provides us with a mathematical proof of global optimisation, which is helpful to verify the result given the same conditions. 

The future direction of maximising the profitability of the cloud broker always has the potential to be improved. One of the few scenarios that we are going to be investigating with our approach is the automation of the risk factors selection. The detail of the effect of each risk factor to the profit is also needed to be examined, namely other anomaly detection in time-series. More realistic scenarios with multiple tiers of instances are to be considered. Additionally, we will be looking at indirect optimisation in more detail to establish a space mapping function between profit and risk parameters.

\section{Conclusion}\label{sect: concl}
We have presented a broker model to simplify the choice of VM pricing schemes for cloud users. Users can enjoy flexibility from a lower pricing scheme, e.g., on-demand instances, while still benefit from the discount from a higher pricing scheme, e.g., reserved instances. The centre of the broker is a risk-analysis based decision-making function to optimise the stock of a VM resource pool. We evaluated the broker is evaluated using a high-frequency real cloud dataset from Alibaba. The results show that the overall profit of the broker is close to the theoretical optimal scenario where user requests can be perfectly predicted.

\balance
\bibliographystyle{IEEEtranN}
\bibliography{mybibliography.bib}
\end{document}